\documentclass[twocolumn]{IEEEtran}
\usepackage{mathptmx}
\usepackage{helvet}
\usepackage{enumitem}
\usepackage{type1cm}

\usepackage{makeidx}         
\usepackage{graphicx}        
\usepackage{multicol}        
\usepackage[bottom]{footmisc}
\usepackage[unicode, pdftex]{hyperref}

\usepackage{amsmath,amssymb,amsfonts, amsthm}
\usepackage{algorithmic}
\usepackage{textcomp}
\makeatletter
\newcommand{\chapterauthor}[1]{%
	{\parindent0pt\vspace*{-25pt}%
		\linespread{1.1}\large\scshape#1%
		\par\nobreak\vspace*{35pt}}
	\@afterheading%
}
\makeatother
\usepackage{amsmath,amssymb, latexsym, amsfonts}
\usepackage{graphicx}
\usepackage{bbm}
\usepackage{xcolor}
\usepackage{tikz, color, soul}
\definecolor{darkgray}{RGB}{64,64,64}
\definecolor{litegray}{RGB}{192,192,192}
\tikzstyle{block}=[draw, rectangle, minimum height=1.5cm, text width=2.5cm, text centered, draw=darkgray]
\tikzstyle{block_large}=[draw, rectangle, minimum height=1.5cm, text width=3cm, text centered, draw=darkgray]
\tikzstyle{line} = [draw, -latex]
\usepackage{newtxmath}
\renewcommand{\(}{\left(}
\renewcommand{\)}{\right)}

\newcommand{\card}[1]{\left|#1\right|}
\newcommand{\eq}{\triangleq}

\newcommand{\I}{{\mathbbm{1}}}

\newcommand{\D}{\mathcal{D}}
\renewcommand{\S}{\mathcal{S}}

\newcommand{\x}{{\textbf{\textit{x}}}}
\newcommand{\y}{{\textbf{\textit{y}}}}

\renewcommand{\u}{{\textbf{\textit{u}}}}
\renewcommand{\v}{{\textbf{\textit{v}}}}

\newcommand{\e}{\varepsilon}

\newcommand{\p}{{\textbf{\textit{p}}}}

\newtheorem{theorem}{Theorem}

\newtheorem{corollary}{Corollary}
\newtheorem{proposition}{Proposition}
\newtheorem{example}{Example}
\newtheorem{definition}{Definition}
\newtheorem{remark}{Remark}

\makeindex             

\begin{document}

	\title{Hypothesis Test for Bounds on the Size of Random Defective Set}
	
	\author{
		Arkadii  D'yachkov, \and Nikita Polyanskii, \and Vladislav Shchukin, \and Ilya Vorobyev
		
		\thanks{The material of this work was presented in part at the 2017 IEEE
			International Symposium on Information Theory. In the present paper, we discuss the extension of the hypothesis testing model considered in the conference paper. Additionally, we prove novel lower and upper bounds on the number of tests presented in Theorem~\ref{thErrorLowerGen} and Theorem~\ref{thUpperBound}, respectively.}

		\thanks{A.  D'yachkov is with the Lomonosov Moscow State University, Moscow 119991, Russia (e-mail: agd-msu@yandex.ru).}
		
		\thanks{N. Polyanskii is with  the Skolkovo Institute of Science and Technology, Moscow 121205, Russia (e-mail: nikitapolyansky@gmail.com).}
		
		\thanks{V.  Shchukin is with the Institute for Information Transmission Problems, Moscow 127051, Russia (e-mail: vpike@mail.ru).}
		
		\thanks{I.  Vorobyev is with the Skolkovo Institute of Science and Technology, Moscow 121205, Russia, and also with the Moscow Institute of Physics and Technology, Dolgoprudny 141701, Russia (e-mail: vorobyev.i.v@yandex.ru).}
		
		\thanks{N. Polyanskii and I.  Vorobyev
			are supported by the Russian Foundation for Basic Research
			under grant No. 18-07-01427~A. }
	}
	
	\maketitle
	
	\begin{abstract}
		The conventional model of disjunctive group testing  assumes that there are several defective elements
	(or defectives) among a large population, and
	a group test yields the positive response if and only if the testing  group
	contains at least one defective element. The basic problem is to find all defectives
	using a minimal possible number of group tests. However, when the number of defectives is unknown
	there arises an additional problem, namely: how to estimate  the random number of  defective elements.
	In this paper, we concentrate on testing the hypothesis $H_0$: the number of defectives $\le s_1$ against the alternative hypothesis $H_1$: the number of defectives $\ge s_2$. We introduce a new decoding algorithm  based on the comparison
	of the number of tests having positive responses with an appropriate fixed threshold. For some asymptotic regimes on $s_1$ and $s_2$, the proposed algorithm is shown to be order-optimal. Additionally, our simulation results verify the advantages of the proposed algorithm such as low complexity and a small error probability compared with known algorithms.   
\end{abstract}
\section{Introduction}
\label{secIntro}
Group testing, also known as Boolean compressed sensing~\cite{atia2012boolean}, is a method for identifying a group of elements with some distinguishable characteristic, frequently referred to as defectives, from a large population.  The main point of the group testing approach is that for a relatively small number of defectives, one can reduce the required number of experiments by testing subgroups of elements rather than all individuals separately. The idea of group testing was introduced by R.~Dorfman~\cite{dor43}.  He proposed
to save on blood tests for infection by grouping individuals and testing the mixture.
The group testing scheme suggested by R. Dorfman is constructed in such a way that
the successive  groups depend on the results of the previous tests.
Such a setting is called adaptive. However, nonadaptive procedures turn out to be useful for practice purposes. A nonadaptive scheme is a series of $N$
group tests that are carried out simultaneously. This is
the essential advantage for the most important applications~\cite{dh00} such as DNA library screening~\cite{dmr00-1}, compressive sensing~\cite{chan2014non}, medical testing~\cite{dor43}, pattern matching algorithms~\cite{clifford2010pattern}.
So, the design and analysis of group
testing algorithms remain an active ongoing area of research.

\subsection{Related work}
The group testing literature may be divided into two categories based on how the number of defectives is modeled. First, let us discuss combinatorial group testing, i.e, the number of defectives, or an upper bound on the number of defectives, is fixed and assumed to be known in advance. Let $t$ be the total  number of elements and $\S$, $\S\subset [t]$,
be an unknown subset of defectives.
The classic group testing problem assumes
that the number of defectives $|\S|$ is upper bounded by some known fixed constant $s$, where the parameter $s$ does not depend on~$t$.  In this regime, the main attention is devoted to disjunctive $s$-codes~\cite{ks64}, which allow finding all the defective elements (a formal definition of disjunctive codes will be given in the next section). 
A specific group testing problem was  discussed in~\cite{sharma2018finding}, where the authors used group tests
to identify a given number of  non-defective items from a large
population containing $s$ defective items.
A more general assumption in combinatorial group testing is that the number of defectives is relatively small, i.e., $\lim \frac{|\S|}{t}  = 0$ as $t \to \infty$.
For example, the regime $|\S| = t^\alpha$, $0 < \alpha < 1$,
was studied  in the recent works~\cite{ald14,ald-scar17,scar17}. Some studies~\cite{bondorf2019sublinear,chan2014non} use random test
designs, and develop computationally efficient algorithms for
identifying defective items from the test outcomes by exploiting the bit-mixing coding and the connection with compressed sensing. 

Many other authors  consider the settings when the number of defectives is unknown. For instance, in the original paper~\cite{dor43}, each element in the population is assumed to be defective
with some fixed probability  $p$. The same model was discussed in the recent papers~\cite{jaggi18,ald18},
where the authors focused on nonadaptive schemes and studied algorithms allowing vanishing error probability as $t\to\infty$.
A more general model in which $p = p(t)$ depends on $t$ was considered in paper~\cite{lev02} of
T.~Berger and V.~Levenshtein. They studied  so-called $2$-stage testing schemes which find all defectives without error.
They proposed to run a fixed number of nonadaptive tests at the first stage and
to test potential candidates individually at the second stage.
For some dependencies $p(t)$, the lower and upper bounds
on the asymptotics of the expected number of tests in the described $2$-stage scheme were obtained in~\cite{lev02,mezard11}.

Now let us refer to the most relevant papers to our work. We first highlight the work of Y. Cheng~\cite{cheng2011efficient}, where the number of defective items can be found with a small error probability using adaptive testing. In~\cite{falahatgar2016estimating}, the authors developed a four-stage adaptive algorithm, which finds the approximate size of the defective set with high probability. One interesting  approach to finding the defective set of unknown size was proposed
by P.~Damaschke and A.S.~Muhammad in~\cite{dam10}. In the beginning, one should
estimate the number of defectives $|\S|$ with the help of  group tests, and then
it remains to use one of the well-known algorithms for finding the defective set $\S$
of the estimated size. We also mention a concept of strict group testing~\cite{damaschke2014strict}, where the searcher must find the defective set when its size is at most $s$ or indicate that the size of the defective set is larger than~$s$. Further details (in particular, the number of tests) for the existing works will be given in Section~\ref{secRelatedRes}.

\subsection{Our contribution}
We consider a model without making any assumptions on the distribution of the number of the subject population, but instead, we focus on modeling the number of defectives using an arbitrary distribution. Our work aims to discuss  testing the hypothesis $H_0$: the  random  number of defectives
is upper bounded by $s_1$ against the alternative hypothesis $H_1$: the  random  number of defectives
is at least $s_2$. The main contributions of this work are two-fold. First, we introduce a new simple testing strategy that uses random tests and compares the number of positive outcomes with an appropriate fixed threshold in order to choose the most likely hypothesis. This low-complexity algorithm can be used for different testing scenarios in practice. Second, we derive closely matching lower and upper bounds on the number of tests required for testing the hypothesis with a small error probability. For instance, our proposed algorithm is shown to be order-optimal when $s_1$ and $s_2$ are asymptotically the same up to a multiplicative constant.
\subsection{Application}
We describe a possible application of our research in  terms of sparse signal models. A signal model is said to be sparse if the number of input variables contributing to the observed outcome (the set of defectives) is relatively small. The study of such models is a new area primarily stimulated by the study of social, computer networks, transportation, power-line. Since most real systems are large and sparse, there are special models developed to understand and analyze them. For consistency of estimation and model prediction, almost all existing methods of variable/feature selection critically depend on the sparsity of models. When this parameter is unknown, we need (at least) to bound the sparsity in the model. Therefore, testing the hypothesis on the size of the defective set can be seen as a preprocessing step for analysis of real-life systems. 

Numerous procedures in biology, medicine and functional genomics require that some bacteria and cells be counted. More frequently, one needs to know only the cell concentration or the concentration of various macromolecules within one or multiple cells in an organism (for example, between $10^3$ and $10^4$ cells per milliliter, or large than $10^{10}$ molecules per cell). Therefore, testing the hypothesis can give crucial information regarding the progress of an infectious disease, or a person's immune system, or the genomic features of different organisms. 
%
\subsection{Outline}
The remainder of this paper is organized as follows.
In Section~\ref{secDefs}, we introduce notation and a hypothesis testing model and give some basic definitions and a conventional algorithm used in group testing. Section~\ref{secRelatedRes} discusses the most relevant results in more detail.
We summarize our results in Section~\ref{secMainResults}. Section~\ref{secModul} is devoted to simulations
of the hypothesis testing problem and comparing different algorithms. The detailed proofs of the main results
will be given in  Section~\ref{secProofs}. Finally, we conclude the paper with Section~\ref{secFinal}.

\section{Problem setup}\label{secDefs}
We introduce several useful notation in Section~\ref{secNot}, describe  a general non-adaptive group testing model in Section~\ref{secGroup}, give the definition of the most conventional codes for group testing in Section~\ref{secDisjunct} and discuss a hypothesis testing model we wish to investigate in Section~\ref{secHyp}.
\subsection{Notation}\label{secNot}
Throughout the paper we adopt the following notation. Let the   symbol $\eq$ denote the equality by definition and
the  symbol $\u \bigvee \v$ denote the disjunctive (Boolean)
sum of binary columns $\u, \v \in \{0, 1\}^N$.
We say that a column $\u$ \textit{covers} a column $\v$ if $\u \bigvee \v = \u$.

For some function $f(x)$ and $g(x)$, we write $f(x)=O(g(x))$ and $f(x)=\Omega(g(x))$ as $x\to\infty$ if there exists some real $x_0$ and $C$ such that $|f(x)|\le C|g(x)|$ and $|f(x)|\ge C|g(x)|$ for $x\ge x_0$, respectively. If both equalities $f(x)=O(g(x))$ and $f(x)=\Omega(g(x))$ hold, then we use notation  $f(x)=\Theta(g(x))$. Finally, we write $f(x)=o(g(x))$ as $x\to\infty$ if $|f(x)|\le \epsilon(x)|g(x)|$ for some function $\epsilon(x)$ such that $\epsilon(x)\to 0$ as $x\to\infty$.

\subsection{Non-adaptive group testing model}\label{secGroup}
In the classical problem of non-adaptive group testing, we describe $N$ tests
as a binary matrix (code) $X\in\{0,1\}^{N\times t}$,
where the $j$th column, denoted by $\x(j)$, corresponds to the $j$th element, and the $i$th row, abbreviated by $\x_i$,
corresponds to the $i$th test. $X$ is often referred to as a testing matrix. Let $x_i(j)$ be the $(i,j)$th element in $X$ and $x_i(j) \eq 1$ if and only if the $j$th element is included into the $i$th testing group.
Let $\S$, $\S \subseteq [t]$, be an arbitrary set
of defective elements of size~$|\S|$. For a code $X$ and a set $\S$,
define the binary \textit{response vector} $\x(\S)$ of length $N$, namely:
$$
\x(\S) \eq
\begin{cases}
\bigvee\limits_{j \in \S} \, \x(j) \quad &\text{if} \quad \S \ne\emptyset, \\
(0, 0, \dots, 0)^T \quad &\text{if} \quad \S = \emptyset.
\end{cases}
$$
The result of a test equals $1$ if at least one defective element is included into the corresponding test
and $0$ otherwise. So the column of test results is equal to the response vector~$\x(\S)$.
\subsection{Disjunctive codes and the conventional decoding algorithm}~\label{secDisjunct}
Now let us give the definition of disjunctive (binary superimposed) codes introduced in~\cite{ks64}.
\begin{definition}
	\label{defDis}
	A binary code $X$ is called a \textit{disjunctive $s$-code} if the disjunctive sum of any $s$-subset
	of columns of $X$ covers those and only those
	columns of $X$ which are the terms of the given disjunctive sum.
\end{definition}
Definition~\ref{defDis} of disjunctive codes
gives the important sufficient condition for identification of any unknown
defective set, namely, one can recover all the defectives based on the response vector if the number of defective elements is at most $s$. In the case of disjunctive $s$-codes, the identification of the unknown set $\S$ is equivalent to searching all columns of
matrix~$X$ covered by~$\x(\S)$,
and its complexity is equal to $\Theta(N t)$ (we run over all $t$ columns in $X$ and compare the covering property of the columns with the response vector of length $N$).
This conventional algorithm of finding defectives is called \textit{COMP
$($Combinatorial Optimal Matching Pursuit$)$}~\cite{ald14}. We refer the reader to~\cite{aldridge2019group} for a survey on decoding algorithms used in group testing.

\subsection{Hypothesis testing model}\label{secHyp}
Given integers $s_1$ and $s_2$ so that $s_1<s_2$, we introduce two hypothesis:
\begin{enumerate}
	\setcounter{enumi}{-1}	
	\item the null hypothesis $\left\{ H_0 \,:\, |\S| \le s_1 \right\}$,
	\item the alternative hypothesis $\left\{ H_1 \,:\, |\S| \ge s_2\right\}$.
\end{enumerate}
In other words, we  want to distinguish reliably two events: the number of defective elements is at most $s_1$ or at least $s_2$. We consider testing the hypothesis using group tests
in the probabilistic model in which
the  random  defective sets of the same size are equiprobable. This assumption is reasonable when there is no prior knowledge on the location of defective elements among the population. In fact, most of the papers in literature discuss such a scenario; e.g., see~\cite{emad2015poisson,d15,dor43,falahatgar2016estimating}.

More accurately, let us define the probability distribution
of the random defective set $\S$ using the vector
$$
\p \eq (p_0, p_1, \dots, p_t),\quad p_k \ge 0,\; k=0,1,\dots,t,
\quad
\sum_{k = 0}^{t} p_k = 1,
$$
in the following manner
\begin{equation}
\label{pd}
\Pr\{\S = \S_0\} \eq \frac{p_{|\S_0|}}{{t \choose |\S_0|}} \quad \text{for any}
\quad \S_0 \subseteq [t].
\end{equation}
Here, $p_i$ corresponds to the probability that there are $i$ defective elements  among the population of size $t$. Since there are ${t \choose i}$ choices to locate $i$ items in the population of size $t$,   the equation~\eqref{pd} does set the probability distribution. Let us provide a few examples of possible probability distributions.
\begin{example}
	In~\cite{emad2015poisson}, the authors consider a truncated Poisson distribution $Pois(t,\lambda)$ and motivate that by the experience in clinical testing. In other words, the probability distribution $\p$ is defined as follows
	$$
	\p = (p_0, p_1, \dots, p_t),\quad p_i = \frac{e^{-\lambda}\lambda^i}{i! m},\quad m=\sum_{i=0}^{t}\frac{e^{-\lambda}\lambda^i}{i!}.
	$$ 
	However, the most popular assumption in group testing is that the vector $\p$ has a Binomial distribution $B(t,p)$, that is
	
	$$
	\p = (p_0, p_1, \dots, p_t),\quad p_i = p^i(1-p)^{t-i}{t \choose i}.
	$$ 
\end{example}
We note that the probability of the null hypothesis and the alternative hypothesis can be expressed with the help of the vector $\p$ as follows
$$
\Pr\{H_0\}=\sum_{i=0}^{s_1}p_i,\quad \Pr\{H_1\}=\sum_{i=s_2}^{t}p_i.
$$
However, the vector $\p$ can be unknown to the searcher. Therefore, we must use some tolerant approach. Let us give some key definitions for hypothesis testing and depict the corresponding model in Fig.~\ref{figH}.
\begin{definition}
An arbitrary map $\D : \{0, 1\}^N \to \{H_0, H_1\}$  is said to be a \textit{decision rule}, which associates a response vector with some hypothesis. Introduce the error probability for the decision rule~$\D$ and the testing matrix~$X$:
\begin{equation}
\label{maxError}
\begin{split}
\e_{s_1}^{s_2}(\p, \D, X) \eq \, \max\, \big\{ \Pr\{ \text{accept } H_1 \big| H_0 \} \, ,
\, \Pr\{ \text{accept } H_0 \big| H_1 \} \,\big\},
\end{split}
\end{equation}
where the probability measure in the conditional probabilities is defined by~(\ref{pd}).
The \textit{universal error probability} is defined by
\begin{equation}
\label{uniError}
\e_{s_1}^{s_2}(\D, X) \eq \, \max\limits_{\p} \, \e_{s_1}^{s_2}(\p, \D, X).
\end{equation}
\end{definition}
Later we  shall omit indices $s_1$ and $s_2$ in notation $\e_{s_1}^{s_2}(\p, \D, X)$ and $\e_{s_1}^{s_2}(\D, X)$, whenever it is clear from the context.
\begin{figure}[t]
	\centering
	\begin{tikzpicture}
	\node[block_large] (c) at (0,-3) {Decision rule $\D:$\\ $\{0,1\}^N\to \{H_0,H_1\}$};
	\node[block_large] (e1) at (0,0) {Testing matrix \\ $[\x_1,\ldots ,\x_N]^T$};
	\node[above] at (0,1) {Encoder};
	\path[draw, dashed, rounded corners] (-2,1) -- (2,1) -- (2,-1) -- (-2,-1) -- cycle;
	\path[draw, dashed, rounded corners] (-2,-2) -- (2,-2) -- (2,-4) -- (-2,-4) -- cycle;
	
	\node[below] at (0,-4) {Decoder};
	\path[line] (-6,0) -- node[below] {$\S\subset[t]$} 	node[above]{Random defective set} (e1);
	\path[line] (c) -- node[below]{$\D(\x(\S))$} node[align = center, above]{Accepting hypothesis}(-6,-3);	
	\path[line] (e1.south) -- node[align=center,left] {Response \\ vector} node[align=center,right] {$\x(\S)$}(c.north);
	\end{tikzpicture}
	\caption{Hypothesis testing model.}
	\label{figH}
\end{figure}
\begin{remark}
Given the testing matrix $X$ and the decision rule $\D$, we search for the worst probability distribution $\p$ and the corresponding maximal error probability in~\eqref{uniError}. In other words, we would like to handle the worst-case scenario which may appear in practice. 
\end{remark}
Let us discuss one important example of hypothesis and a decision rule adopted from the COMP algorithm. Later we will concentrate on this hypothesis and compare this algorithm with one we suggest.
\begin{example}\label{exCOMP}
	Consider the case $s_1=s$ and $s_2=s+1$, that is, we wish to decide whether the number of defective elements is smaller than $s+1$. The COMP algorithm can be used for the hypothesis testing problem in the following way. We say that the COMP decision rule maps the vector $\y\in\{0,1\}^N$ to $H_0$ if the number of columns in $X$ covered by $y$ is at most $s$ and to $H_1$ otherwise. 
	
	For instance, let the testing matrix $X$ and two possible response vectors $\y_1$ and $\y_2$ be as follows
	$$
	X=\begin{pmatrix} 
	0 & 1 & 1 & 0 & 1 \\
	1 & 0 & 0 & 1 & 1\\
	1 & 0 & 1 & 0 & 0 \\
	\end{pmatrix},
	\quad \y_1=\begin{pmatrix} 1 \\ 0\\ 1	\end{pmatrix},
		\quad \y_2=\begin{pmatrix} 1 \\ 1\\ 0	\end{pmatrix}.
	$$
	For $s=2$, the null hypothesis $H_0$ says that the number of defectives is at most $2$, whereas $H_1$ --- the number of defectives is at least $3$. If $\y_1$ is the response vector, then the COMP decision rule accepts $H_0$ (there are two columns, $\x(2)$ and $\x(3)$, covered by $\y_1$). If $\y_2$ is the response vector, then the COMP decision rule accepts $H_1$ (there are three columns, $\x(2)$, $\x(4)$ and $\x(5)$, covered by $\y_2$).
\end{example}
One natural question for testing the hypothesis appears to be as follows. Given the size of population $t$, the error probability level $\epsilon$ and two thresholds $s_1$ and $s_2$, how to minimize the number of tests in a $(N\times t)$ testing matrix $X$ to achieve the universal error probability $\e(\D,X)$ of level at most $\e$. We derive lower and upper bounds on the optimal number of tests required for this problem in Section~\ref{secMainResults}. We carry out Monte-Carlo simulations to estimate the minimal number of tests required for certain hypothesis testing model in Section~\ref{secModul}.
\section{Related Results}
\label{secRelatedRes}
First, we consider testing the hypothesis without error. After that, we outline some state-of-the-art papers on the estimation of the number of defectives with a small error probability. Then we recall some known results for the COMP decision rule used for  hypothesis testing. Finally, we shortly discuss an opportunity to apply the likelihood ratio test for this problem.  
\subsection{Zero-error hypothesis testing}
Let us consider the case $s_1=s$ and $s_2=s+1$. Suppose one wishes to decide whether the number of defectives is smaller than $s+1$ without error, i.e., $\epsilon(\D,X)=0$ for some decision rule $\D$ and testing matrix $X$.  The problem of optimal zero-error nonadaptive hypothesis testing is reduced to
the problem of optimal disjunctive codes with the help of the following statement. This proposition turns out to be a group testing folklore result.
\begin{proposition}
\label{propDisj}
A code  $X$ is a disjunctive $s$-code if and only if
for any probability distribution $\p$ with positive  components
$p_s > 0 \text{ and } p_{s + 1} > 0$, there exists a decision rule $\D$
such that the error probability $\e(\p, \D, X) = 0$.
\end{proposition}
\begin{proof}
If $X$ is a disjunctive $s$-code, then obviously the COMP decision rule allows us to check the hypothesis $H_0$ without error (see Example~\ref{exCOMP}).
The converse result can be proved by contradiction. Indeed, if the matrix $X$ is not a disjunctive $s$-code, then there exists
 a set $\S \subseteq [t]$ of size $|\S| = s$,
and a number $j \in [t] \setminus \S$ such that $\x(\S) = \x(\S \cup \{j\})$.
So, for any decision rule we cannot distinguish the set $\S$ of size $s$ from the set $\S \cup \{j\}$ of size~$s + 1$.
\end{proof}
The best known practical constructions of disjunctive  $s$-codes are
based on shortened Reed Solomon codes. These constructions presented
in~\cite{dmr00-1} essentially  extend  optimal
and suboptimal ones
suggested in~\cite{ks64}.

Recall some results 
for optimal disjunctive $s$-codes. Denote by $N(t,s)$ the minimal number of rows for disjunctive\quad $s$-codes with $t$ columns.
The best known lower and upper bounds on $N(t,s)$ are presented in~\cite{dr82} and~\cite{drr89}, respectively.
These bounds are written in the complex form, but for fixed integer $s$, the asymptotics are as follows
\begin{align*}
N(t,s) =&\, O\left(s^2\ln t\right)  \quad \text{as }t \to \infty,\\
N(t,s) =&\, \Omega\left(\frac{s^2}{\ln s}\ln t\right)  \quad \text{as }t \to \infty.
\end{align*}
\subsection{Estimating the number of defectives with a small error probability}
Further we will consider the case of positive error probability. In~\cite{dam10} the authors present a randomized algorithm that uses
$g(\epsilon, c) \log_2 t$ nonadaptive tests (here, $g(\epsilon, c)$ is a some function which depends only on $c$ and $\epsilon$) and produces some statistic $\hat s$
which satisfies the following properties: probability
$\Pr\{\hat s < |\S|\}$ is upper bounded by a small parameter $\epsilon \ll 1$
and the expected value of~$\hat s / |\S|$ is upper bounded by a number~$c > 1$.
Note that this result is universal, i.e., it does not depend on the distribution
of the defective set. In~\cite{orl16} the authors construct an adaptive randomized algorithm
which uses at most $2 \log_2 \log_2 |\S| + O(\frac{1}{\delta^2} \log_2 \frac{1}{\epsilon})$
adaptive tests and estimates $|\S|$ up to a multiplicative factor of $1 \pm \delta$ with error probability $\le \epsilon$.
Also there is a converse result in~\cite{orl16} which states
the necessity of $(1 - \epsilon) \log_2 \log_2 |\S| - 1$ tests on average.
\subsection{COMP decision rule for hypothesis testing}
Let us consider the case $s_1=s$ and $s_2=s+1$ again. The first thought which comes to mind from Proposition~\ref{propDisj} is that in order to solve the hypothesis testing problem one may use the COMP decision rule.
Note that this rule always accepts $H_1$ if it holds, i.e., $\Pr(H_0|H_1)=0$.
Moreover, it is not difficult to obtain that the maximum of the error probability $\e(\p, \text{COMP}, X)$
in (\ref{uniError}) is attained at any vector $\p$ such that $0< p_s < 1$ and $p_k = 0$ for $\forall \, k < s$, e.g., one can take $p_s=p_{s+1}=1/2$.
The reader may refer to Proposition~\ref{lemWorstProb} about a similar statement which
is given in Section~\ref{secMainResults} and proved in Section~\ref{secProofLemWorstProb}.
That is why the universal error probability $\e_s(\text{COMP}, X)$ equals the probability that
an $s$-subset of columns of $X$ covers an external column. But this probability is exactly
the error probability for almost disjunctive $s$-codes~\cite{d15}.
The properties of the universal error probability $\e(\text{COMP}, X)$ obtained
in~\cite{d15} are presented below as Propositions~\ref{thCapacityUpper}
and~\ref{thCapacityLower}.
\begin{proposition}[Follows from Theorem $2$ in~\cite{d15}] 
\label{thCapacityUpper} Let $\epsilon>0$ be a fixed real scalar, and $s,\, t$ be integers such that $s\le t$.
If $X$ is an arbitrary code of size $t$ such that  the universal error probability  for the COMP decision rule and the testing matrix $X$ is less than $\e$, then length of the code $X$ must be $N=\Omega\left(s\ln \left(t/s\right)\right)$ as $t \to \infty$.
\end{proposition}
\begin{remark}
In other words, Proposition~\ref{thCapacityUpper} says that to guarantee a vanishing error probability,  the number of tests, accompanied with the COMP decision rule, should be linear  with $\ln (t/s)$, the logarithm of the ratio between the size of the population and the number of defectives.  It will be shown in Theorem~\ref{thUpperBound} that for the hypothesis testing problem, the optimal number of tests does not depend asymptotically on the size $t$.
\end{remark}
In~\cite{d15}, using the probabilistic method we  established the existence result on the almost disjunctive codes. We interpret that results in terms of hypothesis testing.
\begin{proposition}[Follows from Theorem $4$ in~\cite{d15}]
\label{thCapacityLower}\quad
Given real scalar $\epsilon>0$ and fixed integer $s>0$, there exists a $(t\times N)$ testing matrix $X$ with $N= O(s \ln t)$ tests such that the COMP decision rule provides the universal error probability less than $\epsilon$. 
\end{proposition}
\begin{remark}
	Given fixed integer $s>0$, by Propositions~\ref{thCapacityUpper} and~\ref{thCapacityLower}, the COMP decision rule needs $\Theta(s\ln t)$ tests to  distinguish the null hypothesis $H_0$: the number of defective elements is at most $s$ from the alternative hypothesis $H_1$: the number of defective elements is at least $s+1$.
\end{remark}
\subsection{Alternative decision rule} \label{Subsect:Alt}
For the case $s_2=s_1+1$, one may try to use the generalized likelihood ratio test. Such a test has critical region $R:= \{\y: \lambda(\y) \le a\}$,
where
$$
\lambda(\y):=\frac{\max_{s\le s_1}\Pr\{\x(\S)=\y\big| |\S|=s\}}{\max_{s\in[t]}\Pr\{\x(\S)=\y\big| |\S|=s\}}
$$
is the generalized likelihood ratio and parameter $a$ is chosen to satisfy requirements on $\Pr\{H_1|H_0\}$. In other words, we accept $H_1$ when $\lambda(\y)\in R$. However, the decoding complexity of this rule is exponential with the number of items.
\section{Main Results}
\label{secMainResults}
Note that our hypothesis testing problem is very different from the finding the defectives
because there are only two answers: $H_0$ or $H_1$. However, in the zero-error case, due to Proposition~\ref{propDisj} testing the hypothesis
requires nearly the same number of group tests as the searching the defectives does.
Our main results are devoted to hypothesis group testing in the case of a small error probability. First, we introduce a low-complexity weight decision rule
which turns out to be better than the COMP decision rule in terms of the optimal number of tests. Then Proposition~\ref{lemWorstProb} shows that the worst probability distribution $\p$ in~\eqref{uniError} for the weight decision rule should be concentrated in only two points $p_{s_1}$ and~$p_{s_2}$. Finally we obtain lower and upper bounds on the optimal number of tests for the hypothesis testing problem in Section~\ref{secLower} and Section~\ref{secUpper}, respectively. All the proofs will be given in Section~\ref{secProofs}.
\subsection{Weight decision rule}
Fix an arbitrary parameter $\tau$, $0 < \tau < 1$, and
introduce a \textit{$\tau$-weight decision rule ($\tau$-WDR)}
\begin{equation}
\label{wdr}
\begin{cases}
\text{accept $\left\{ H_0 \,:\, |\S| \le s_1 \right\}$} & \text{if $|\x(\S)| \le \tau N$}, \\
\text{accept $\left\{ H_1 \,:\, |\S| > s_2 \right\}$} & \text{if $|\x(\S)| > \tau N$}. \\
\end{cases}
\end{equation}

\begin{remark}
We note that the $\tau$-weight decision rule is related to some model of specific disjunctive $s$-codes considered in~\cite{b15}. The authors of that paper supply a disjunctive $s$-code with a weaker additional condition:
the weight $|\x(\S)|$ of the response vector for any subset $\S$, $\S \subseteq [t]$, $|\S| \le s$, is at most $T$.
Such a group testing model is motivated
by a risk for the safety of the persons who perform tests, in some contexts,
when the number of positive test results is too large.
\end{remark}
\subsection{Worst-case probability distribution for the weight decision rule}
We study only the universal error probability for the weight decision rule,
and the following statement proved in Section~\ref{secProofLemWorstProb} determines
the worst probability distribution in~\eqref{uniError}.
\begin{proposition}
\label{lemWorstProb}
For any integers $s_1,s_2<t/2$, real scalar $\tau,\,0<\tau <1,$ and the testing $(N\times t)$ matrix $X$, the maximum of the error probability $\e(\p, \text{$\tau$-WDR}, X)$
in (\ref{uniError}) is attained at any $\p$ such that $p_{s_1} > 0$, $p_{s_2} > 0$ and $p_{s_1} + p_{s_2} = 1$.
\end{proposition}

\subsection{Lower bound on the number of tests}\label{secLower}
The next converse theorem derives the lower bound on the error probability
for the worst distribution from Proposition~\ref{lemWorstProb} and any decision rule.
\begin{theorem}
\label{thErrorLowerGen}
Let the distribution $\p$ be such that $p_{s_1} > 0$, $p_{s_2} > 0$ and $p_{s_1} + p_{s_2} = 1$.
For any decision rule $\D$ and any testing $(N\times t)$ matrix $X$ with $N\le s_1 \log_2( t/s_2)$,
the error probability is lower bounded by
$$
\e(\p, \D, X) \ge\frac{\left(2^{-\frac{N}{s_1}} t - s_1\right)\ldots\left(2^{-\frac{N}{s_1}} t - s_2+1\right)}{2(t-s_1)\ldots(t-s_2+1)}.
$$
\end{theorem}
This lower bound consequently leads to the bound on the minimal number of tests for optimal testing strategy. We focus on two special cases: when two thresholds $s_1$ and $s_2$ are the same up to a multiplicative factor or an additive factor.
\begin{corollary}\label{corLowerBound}
Let $s_1<s_2$ and $s_2 = o(t)$. For any $(N\times t)$-code $X$ 
and any decision rule $\D$ with the universal error probability $\e(\D, X) \le \epsilon$, the number of non-adaptive group tests is lower bounded as follows
$$
N\ge\min\left\{s_1\log_2(t/s_2),\,\frac{s_1}{s_2-s_1}\log_2\frac{1}{\e}(1+o(1))\right\} \quad\text{as }\epsilon\to0.
$$
$(a)$ Given the real number $\alpha>1$, let $s_1=s$ and $s_2=\alpha s$. Then  
$$
N\ge\min\left\{s_1\log_2(t/s_2),\,\frac{1}{\alpha -1}\log_2\frac{1}{\e}(1+o(1))\right\} \quad\text{as }\epsilon\to0.
$$
$(b)$ Given the integer $c=o(s)$, let $s_1=s$ and $s_2=s+c$. Then  
$$
N\ge\min\left\{s_1\log_2(t/s_2),\,\frac{s}{c}\log_2\frac{1}{\e}(1+o(1))\right\} \quad\text{as }\epsilon\to0.
$$
\end{corollary}
\subsection{Upper bound on the number of tests}\label{secUpper}
The next statement establishes an upper bound on the optimal number of tests. The testing matrix is constructed probabilistically, with the entries of the matrix being independent and Bernoulli distributed. We make use of the fact that a Binomial distribution is concentrated around its mean, with exponentially small tail. 
\begin{theorem}\label{thUpperBound}
	Given two thresholds $s_1$ and $s_2$, real scalar $\epsilon>0$ and integer $t>0$, for any integer $N\ge N_0$, there exists a testing $(N\times t)$-matrix $X$ such that the $\tau$-weight decision rule provides the universal error probability $\e(X,\tau\text{-WDR})$ for testing the hypothesis $H_0$ against $H_1$ less than $\epsilon$, where
	$$
	N_0\eq\frac{1}{2\delta^2}\ln\frac{1}{\epsilon},\quad \tau\eq1-(1-p)^{s_1}/2-(1-p)^{s_2}/2,
	$$
	$$
	\delta\eq (1-p)^{s_1}/2-(1-p)^{s_2}/2,\quad p \eq \frac{\ln s_2 - \ln s_1}{s_2-s_1}.
	$$
$(a)$ Given the real number $\alpha>1$, let $s_1=s$ and $s_2=\alpha s$. Then  
$$
N_0=\frac{2\alpha^{\frac{2\alpha}{\alpha-1}}}{(\alpha-1)^2}\ln\frac{1}{\epsilon}(1+o(1)) \quad\text{as }s\to\infty.
$$
$(b)$ Given the integer $c=o(s)$, let $s_1=s$ and $s_2=s+c$. Then  
$$
N_0=\frac{2e^2s^2}{c^2}\ln\frac{1}{\e}(1+o(1)) \quad\text{as }s\to\infty.
$$
\end{theorem}
\begin{remark}
	Independence from the number of elements $t$ is crucial in Theorem~\ref{thUpperBound}.
	In other words, we can construct a sequence of $(N \times t(N))$ matrices
	with exponentially decreasing error probability for any function $t(N)$. Due to Proposition~\ref{thCapacityUpper} the number of tests should be linear with $\ln t$ whenever the error probability for the COMP decision rule is vanishing.
	Therefore, the weight decision rule has a significant advantage
	over the COMP decision rule when the size of the population is large.
\end{remark}
Let us also mention a stronger bound from the conference paper~\cite{d17_isit} than one given in Theorem~\ref{thUpperBound}. This result has more narrow applications and is related to distinguishing the null hypothesis $H_0$: the number of defective elements is at most $s$ from the alternative hypothesis $H_1$: the number of defective elements is at least $s+1$. The proof of the following statement is quite technical and is based on the probabilistic method. We generate random codes with a fixed Hamming weight and derive an upper bound on the number of tests for hypothesis testing.
\begin{theorem}[Theorem $3$ in~\cite{d17_isit}]
	\label{thErrorUpperOld}
	Given the real scalar $\epsilon>0$ and the integer $t$, for any integer $N\ge N_0$, there exists a testing $(N\times t)$-matrix $X$ with fixed weight $wN$ such that the $\tau$-weight decision rule provides the universal error probability $\e(X,\tau\text{-WDR})$ for testing the hypothesis $H_0$ against $H_1$ less than $\epsilon$, where
	\begin{equation*}
	\label{lowerEAs}
N_0 \eq 4s^2 \ln \frac{1}{\e}(1+o(1))\quad \tau \eq \frac{c}{s},
$$
$$
w\eq  \frac{c}{s^2},\quad c = o(s) \quad\text{as }s\to\infty. 
	\end{equation*}
\end{theorem}
\begin{remark}\label{remSim}
	By Theorem~\ref{thUpperBound} Claim $(b)$, we need $2e^2 s^2\ln\frac{1}{\e}\approx 14.8s^2\ln\frac{1}{\e}$ tests to provide the error probability less than $\e$ when testing whether the number of defectives is less than $s+1$, whereas Theorem~\ref{thErrorUpperOld} guarantees that $4s^2\ln\frac{1}{\e}$ tests are enough. Another advantage of Theorem~\ref{thErrorUpperOld} is that the testing matrix has a low density. Indeed, the relative weight of columns turns out to be of order $1/s^2$ (versus $1/s$ in Theorem~\ref{thUpperBound}). Therefore, for practical applications, we recommend to generate a constant-weight testing matrix with the relative Hamming weight of order $1/s^2$ to test the hypothesis that the number of defectives is at most $s$ against the hypothesis that the number of defectives is at least $s+1$.
\end{remark}
\section{Simulation}
\label{secModul}
In this section, we first compare the weight decision rule and the COMP decision rule in a specific setting. Second, we estimate more carefully the minimum number of tests required for the weight decision rule to guarantee small error probability in the setting when the number of items can be arbitrary large and (i) $s_2-s_1= \Theta(1)$, (ii) $s_2-s_1= \Theta(\sqrt{s_1})$, (iii) $s_2-s_1= \Theta({s_1})$.
\subsection{Comparison between the COMP decision rule and the weight decision rule}
Now we carry out Monte Carlo simulations to find the minimal number of tests required for the COMP decision rule and the weight decision rule to guarantee the universal error probability  less than the level $\epsilon = 0.01$. 

\begin{figure}
	\centering
	\includegraphics[width=1\linewidth]{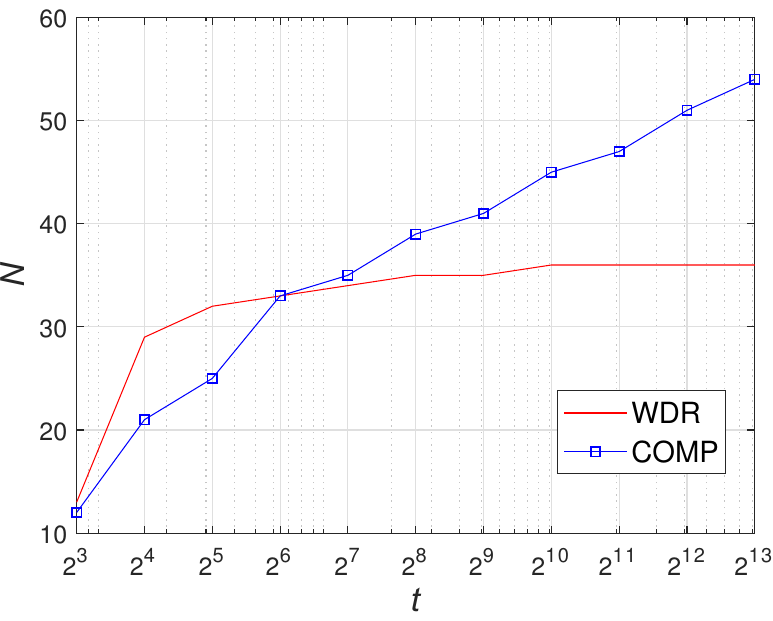}
	\caption{``Optimal'' number of tests for the COMP decision rule and the weight decision rule estimated by the Monte-Carlo simulation method.}
	\label{fig:compvswdr}
\end{figure}

We test the hypothesis $H_0$: the number of defectives is at most $2$, against the hypothesis $H_1$: the number of defectives is at least $3$. To this end, for all $t\in\{8,16,32,\ldots, 8192\}$, following Remark~\ref{remSim}, we generate $(N \times t)$ binary matrices of some constant weight $w$ with minimal possible $N$ so that the universal error probability is less than $\e$. We generate matrices with different $w\in\{1,\ldots,N/2\}$ and repeat the procedure $100$ times for any given $w$ and $N$ to find a matrix with the universal error probability less than $\e$. To estimate the universal error probability of some testing matrix and some decision rule, we employ the Monte Carlo  method, namely, subsets $\S\subset[t]$, of size $2$ and $3$ are chosen randomly $10000$ times and the corresponding error probability is estimated using~\eqref{Bk}-\eqref{maxerror2}. We depict our results in Fig.~\ref{fig:compvswdr}, where the $x$-axis corresponds to different $t$, number of items,  and the $y$-axis corresponds to ``optimal'' $N$, number of tests. Additionally, a base-$2$ logarithmic scale is used for the $x$-axis. One can easily see the series of  numbers of tests for the weight decision rule converges to some level as $t$ is growing, whereas the number of tests for the COMP decision rule is linear with $\ln t$, the logarithm of the number of items.

The simulation results verify the advantage of the WDR
over the COMP decision rule in pursuing the minimal possible number of tests. Also, we recall that the decoding complexity of the WDR decision rule is $\Theta(N)$, whereas the complexity of the COMP decision rule is significantly higher, namely, $\Theta(Nt)$. 
\subsection{Weight decision rule when the number of items is large}
Now we consider the case when the number of items, denoted by $t$, can be arbitrary large. We apply the weight decision rule to test the hypothesis $H_0$: the number of defectives is at most $s$, against the hypothesis $H_1$: the number of defectives is at least (i) $s+10$, (ii) $s+\sqrt{s}$, (iii) $1.25s$. Let the universal error probability be at most $\epsilon=0.001$. By Corollary~\ref{corLowerBound} and Theorem~\ref{thUpperBound} we know that the optimal number of tests in these regimes as $s\to\infty$ can be bounded as follows 
\begin{itemize}
\item[(i)] $9.9 s=\frac{s}{c}\log_2\frac{1}{\e}\lesssim N\lesssim \frac{2e^2s^2}{c^2}\ln\frac{1}{\e}=1.02s^2$;
\item[(ii)] $9.9 \sqrt{s}=\frac{s}{c}\log_2\frac{1}{\e}\lesssim  N\lesssim\frac{2e^2s^2}{c^2}\ln\frac{1}{\e}=102s$;
\item[(iii)] $39.9=\frac{1}{\alpha -1}\log_2\frac{1}{\e}\le N\lesssim \frac{2\alpha^{\frac{2\alpha}{\alpha-1}}}{(\alpha-1)^2}\ln\frac{1}{\epsilon}=2058.7$.
\end{itemize}

\begin{figure}
	\centering
	\includegraphics[width=1\linewidth]{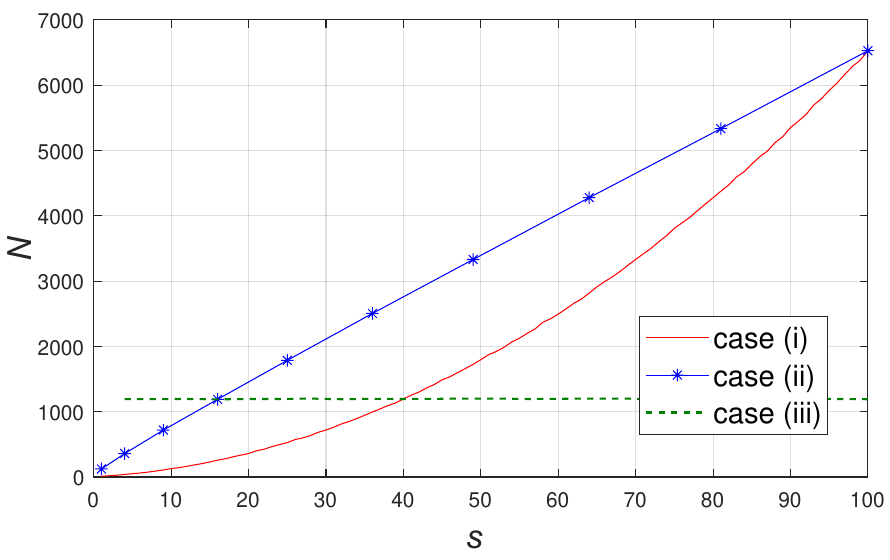}
	\caption{Sufficient number of tests in a random testing matrix}
	\label{fig:threeGraphs}
\end{figure}

For i) $s\in\{1,2,3,4,\ldots,100\}$, ii) $s\in\{1,4,9,16,\ldots,100\}$, iii) $s\in\{4,8,12,16,\ldots,100\}$, we estimate the minimal number of tests in a random testing matrix with Bernoulli entries more carefully than in the proof of Theorem~\ref{thUpperBound}. More precisely,  we compute probabilities~\eqref{EstimateProb1}-\eqref{EstimateProb2} directly instead of applying Hoeffding's inequalities. The results are depicted in Fig.~\ref{fig:threeGraphs}, where the $x$-axis corresponds to different $s$, thresholds in the testing model, and the $y$-axis corresponds to the number of tests. 
\section{Proofs of Main Results}
\label{secProofs}

\subsection{Proof of Proposition~\ref{lemWorstProb}}
\label{secProofLemWorstProb}
\begin{proof}
For an arbitrary binary $(N \times t)$-matrix $X$ and parameters $s_1,\,s_2$ and $T \eq \lfloor \tau N \rfloor$, introduce the sets $B^i_k(T, X)$,
$i = 1, 2$, $k = 0, 1, \dots, t$, of $k$-subsets of set $[t]$ as follows:
\begin{equation}
\label{Bk}
\begin{split}
B^1_k(T, X) &\eq \, \left\{ \, \S \,:\, \S \subseteq [t], \; |\S| = k, \; |\x(\S)| \ge T + 1 \right\},\\
B^2_k(T, X) &\eq \, \left\{ \, \S \,:\, \S \subseteq [t], \; |\S| = k, \; |\x(\S)| \le T \right \}.
\end{split}
\end{equation}
Then the error probability for the $\tau$-weight decision rule is represented by
\begin{multline}
\label{maxerror2}
\e(\p, \text{$\tau$-WDR}, X)
\\
\eq \max \Bigg\{ \sum_{k = 0}^{s_1} \frac{p_k}{\sum\limits_{l = 0}^{s_1} p_l} \frac{\card{B^1_k(T, X)}}{{t \choose k}},
\sum_{k = s_2}^{t} \frac{p_k}{\sum\limits_{l = s_2}^t p_l} \frac{\card{B^2_k(T, X)}}{{t \choose k}} \Bigg\}.
\end{multline}

For any $k<t$, $\S\in B^1_k(T, X)$, and $j\in[t]\setminus \S$,  one can construct a set $\S'\eq\S\cup \{j\}$
belonging to $B^1_{k + 1}(T, X)$. Since the number of choices for $j$ is $(t-k)$ and there exist at most $(k+1)$ ways leading to the same set $\S'\in B^1_{k + 1}(T, X)$, we obtain the following inequality:
$$
|B^1_{k+1}(T, X)| \ge \frac{t-k}{k+1} |B^1_k(T, X)|.
$$
Applying such inequality $(s_1-k)$ times, we obtain for any $k<s_1$
$$
|B^1_{s_1}(T,X)|\ge \frac{s_1\ldots (k+1)}{(t-s_1+1)\ldots(t-k)}|B^1_k(T,X)|.
$$
Equivalently, we have 
\begin{equation}\label{toS1}
\frac{|B^1_k(T, X)|}{{t \choose k}}\le \frac{|B^1_{s_1}(T, X)|}{{t \choose s_1}}.
\end{equation}
Similarly, one can construct set $\S \in B^2_k(T, X)$ by removing from any set $\S' \in B^2_{k + 1}(T, X)$
any index $j \in \S'$, and at most $(t - k)$ such different pairs $(\S', j)$
may construct the same $\S \eq \S' \setminus \{j\}$. Therefore
$$
|B^2_{k}(T, X)| \ge \frac{k+1}{t-k} |B^2_{k+1}(T, X)|.
$$
Thus, for any $k>s_2$, we have
\begin{equation}\label{toS2}
\frac{|B^2_k(T, X)|}{{t \choose k}}\le \frac{|B^2_{s_2}(T, X)|}{{t \choose s_2}}.
\end{equation}
Definition~(\ref{maxerror2}) and  inequalities~\eqref{toS1}-\eqref{toS2} yield
\begin{equation}
\label{maxerrorUpper}
\e(\p, \text{$\tau$-WDR}, X) \le \max \left\{ \frac{\card{B^1_{s_1}(T, X)}}{{t \choose s_1}}, \frac{\card{B^2_{s_2}(T, X)}}{{t \choose s_2}} \right\},
\end{equation}
and equality holds in~(\ref{maxerrorUpper}) for any distribution with the properties:
$p_{s_1} > 0$, $p_{s_2} > 0$, and $p_j = 0$ for $ j \in [t]\setminus\{s_1, s_2\}$.
In particular, it means that for $\tau$-WDR the definition of the universal error probability~(\ref{uniError})
is equivalent to the right-hand side of~(\ref{maxerrorUpper}).
\end{proof}
\subsection{Proof of Theorem~\ref{thErrorLowerGen}}
\label{secProofThErrorLower}
\begin{proof}
	Let an $(N \times t)$-matrix $X$ be a testing matrix,
	$\D: \{0, 1\}^N \to \{H_0, H_1\}$ be a decision rule
	and $\p$ be a distribution so that $p_{s_1} > 0$, $p_{s_2} > 0$ and $p_j = 0$ for any $j \in [t]\setminus\{s_1, s_2\}$.
	Obviously, the maximal error probability~(\ref{maxError}) is bounded below
	by the half of the sum:
	\begin{equation}
	\label{maxErrorSum}
	\e_s(\p, \D, X) \ge \frac{1}{2} \( \Pr\{ \text{accept } H_1 \big| H_0 \} + \Pr\{ \text{accept } H_0 \big| H_1 \} \).
	\end{equation}
	Denote the number of $k$-subsets of columns with the response vector $\y$ by $n_k(\y, X)$, i.e.,
	$$
	n_k(\y, X) \eq \left| \{ \S : |\S| = k, \; \x(\S) = \y \} \right|.
	$$
	Since the distribution of $\p$ is concentrated in only two coordinates $p_{s_1}$ and $p_{s_2}$, we can rewrite the right-hand side of~\eqref{maxErrorSum}  as follows
	\begin{multline}
	\label{maxErrorBigSum}
	\e(\p, \D, X) \ge \frac{1}{2} \sum_{\y \in \{0, 1\}^N} \(
	\frac{n_{s_1}(\y, X)}{{t \choose s_1}} \I \{\D(\y) \not= H_0\} \right. \\ \left.+
	\frac{n_{s_2}(\y, X)}{{t \choose s_2}} \I \{\D(\y) \not= H_1\} \).
	\end{multline}
	One of the two indicators $\I \{\D(\y) \not= H_0\}$, $\I \{\D(\y) \not= H_1\}$
	equals $0$ and the other one equals $1$.  Therefore
	\begin{equation}
	\label{maxErrorMinSum}
	\e(\p, \D, X) \ge \frac{1}{2} \sum_{\y \in \{0, 1\}^N}
	\min \left\{ \frac{n_{s_1}(\y, X)}{{t \choose s_1}} ,
	\frac{n_{s_2}(\y, X)}{{t \choose s_2}} \right\}.
	\end{equation}
	Further we consider only such $\y$'s that $n_{s_1}(\y, X) > 0$.
	It is obvious that for other $\y$'s the minimum in the sum~(\ref{maxErrorMinSum}) equals $0$.
	Denote the relative number of $s_1$-subsets with the response vector $\y$ by $\beta_\y$, i.e.,
	$\beta_\y \eq n_{s_1}(\y, X) / {t \choose s_1}$, and note that
	\begin{multline*}
	n_{s_1}(\y, X) = \beta_\y {t \choose s_1} = \beta_\y \frac{(t - s_1 + 1) \ldots t}{s_1!}\\
	\ge \frac{(\beta_\y^{\frac{1}{s_1}} t - s_1 + 1) \ldots (\beta_\y^{\frac{1}{s_1}} t)}{s_1!}
	= {\beta_\y^{\frac{1}{s_1}} t \choose s_1}
	\end{multline*}
	because $0 < \beta_\y \le 1$. By $\S_\y$ denote a set of all column's indices which are included
	into some $s_1$-subset $\S$ so that the response vector $\x(\S)=\y$, i.e.,
	$$
	\S_\y = \bigcup_{\S : \x(\S) = \y} \S,
	$$
	and suppose that $|\S_\y| = s_1 + L$ for some integer $L$. The previous inequality and this assumption lead to
	$$
	{\beta_\y^{\frac{1}{s_1}} t \choose s_1} \le n_{s_1}(\y, X) \le {s_1 + L \choose s_1}.
	$$
	It follows that $L \ge \beta_\y^{\frac{1}{s_1}} t - s_1$ (the right-hand side can be replaced by the ceiling).
	Given $\S$, $|\S|=s_1$, with the response vector $\x(\S) = y$, and $(s_2-s_1)$ distinct indices $j_1,\ldots,j_{s_2-s_1}\in \S_\y \setminus \S$, one
	one can construct the $s_2$-subset $\S'=\S\cup\{j_1,\ldots, j_{s_2-s_1}\}$ with the same response vector $\x(\S')=\y$.
	Moreover, any $\S'$ can be constructed in at most ${s_2 \choose s_1}$ such ways. Hence
	\begin{multline}
	\label{nyLowerBound}
	n_{s_2}(\y, X) \ge  n_{s_1}(\y, X)\frac{{L \choose s_2 - s_1}}{{s_2 \choose s_1}}\ge \\\beta_\y {t \choose s_1}  \frac{\left(\beta_\y^{\frac{1}{s_1}} t - s_1\right)\ldots\left(\beta_\y^{\frac{1}{s_1}} t - s_2+1\right)}{s_2(s_2-1)\ldots(s_1+1)} \\ =
		\beta_\y {t \choose s_2}  \frac{\left(\beta_\y^{\frac{1}{s_1}} t - s_1\right)\ldots\left(\beta_\y^{\frac{1}{s_1}} t - s_2+1\right)}{(t-s_1)\ldots(t-s_2+1)}.
	\end{multline}
	Therefore, the minimum in the sum~(\ref{maxErrorMinSum}), that is,
	$$
\min \left\{ \frac{n_{s_1}(\y, X)}{{t \choose s_1}} ,
\frac{n_{s_2}(\y, X)}{{t \choose s_2}} \right\}
	$$
	 is bounded below by
	$$
	\min\left\{\beta_\y,\, \beta_\y  \frac{\left(\beta_\y^{\frac{1}{s_1}} t - s_1\right)\ldots\left(\beta_\y^{\frac{1}{s_1}} t - s_2+1\right)}{(t-s_1)\ldots(t-s_2+1)} \right\}.
	$$
	Recall that $0 < \beta_\y \le 1$. Thus, we have $\frac{\beta_\y^{\frac{1}{s_1}} t - c}{t-c}\le 1$ for any real $c,\, c<t$, and the above minimum is attained at the second argument. It follows
	\begin{multline*}
	\min \left\{ \frac{n_{s_1}(\y, X)}{{t \choose s_1}} ,
	\frac{n_{s_2}(\y, X)}{{t \choose s_2}} \right\} \\ \ge \beta_\y  \frac{\left(\beta_\y^{\frac{1}{s_1}} t - s_1\right)\ldots\left(\beta_\y^{\frac{1}{s_1}} t - s_2+1\right)}{(t-s_1)\ldots(t-s_2+1)}.
	\end{multline*}
	Finally, we conclude
	\begin{multline*}
	\e(\p, \D, X) \ge \\
	\frac{1}{2} \sum_{\y \in \{0, 1\}^N : \beta_\y > 0}
	\beta_\y  \frac{\left(\beta_\y^{\frac{1}{s_1}} t - s_1\right)\ldots\left(\beta_\y^{\frac{1}{s_1}} t - s_2+1\right)}{(t-s_1)\ldots(t-s_2+1)}
	 \\= \frac{1}{2}\sum_{\y \in \{0, 1\}^N : \beta_\y > 0} \beta_\y f(\beta_\y^{-1}) \overset{(a)}{\ge} \frac{1}{2}f\( \sum_{\y \in \{0, 1\}^N : \beta_\y > 0} \beta_\y \cdot \beta_\y^{-1} \)\\
	 \overset{(b)}{\ge} \frac{\left(2^{-\frac{N}{s_1}} t - s_1\right)\ldots\left(2^{-\frac{N}{s_1}} t - s_2+1\right)}{2(t-s_1)\ldots(t-s_2+1)},
	\end{multline*}
	where the function $f(x)$ is defined by
	$$
	f(x) \eq  \frac{\left(x^{-\frac{1}{s_1}} t - s_1\right)\ldots\left(x^{-\frac{1}{s_1}} t - s_2+1\right)}{(t-s_1)\ldots(t-s_2+1)}.
	$$
We use $i)$ the property $\sum_{\y \in \{0, 1\}^N} \beta_\y = 1$,  $ii)$ the inequality $f(x)>0$ for all $x\in [1, 2^N]$ and  $iii)$ Jensen's inequality for the $\cup$-convex function $f(x)$  in the interval $x\in [1, 2^N]$ to prove the inequality $(a)$. To obtain the inequality $(b)$, we observe $i)$ the number of distinct $\y$'s in the sum  is at most $2^N$ or
$$
\sum_{\y \in \{0, 1\}^N : \beta_\y > 0} \beta_\y \cdot \beta_\y^{-1} \le  2^N,
$$
and $ii)$ the function $f(x)$ is decreasing in the interval $[1, 2^N]$.
This completes the proof.
\end{proof}
\subsection{Proof of Corollary~\ref{corLowerBound}}
\begin{proof}
By Theorem~\ref{thErrorLowerGen}, we know that the universal error probability of any testing matrix $X$ and any decision rule $\D$ is bounded below by
$$
\e(\D, X) \ge\frac{\left(2^{-\frac{N}{s_1}} t - s_1\right)\ldots\left(2^{-\frac{N}{s_1}} t - s_2+1\right)}{2(t-s_1)\ldots(t-s_2+1)}.
$$
Since the universal error probability should be at most $\epsilon$ we obtain
$$
\epsilon \ge \frac{\left(2^{-\frac{N}{s_1}} t - s_1\right)\ldots\left(2^{-\frac{N}{s_1}} t - s_2+1\right)}{2(t-s_1)\ldots(t-s_2+1)}.
$$
Recall $s_2 = o(t)$. Suppose that all the factors in the right-hand side are positive. By taking the binary logarithm, we get
$$
\log_2 \epsilon \ge -\frac{(s_2-s_1)N}{s_1}(1+o(1)).
$$
Therefore, 
$$
N\ge\frac{s_1}{s_2-s_1}\log_2 \frac{1}{\epsilon} (1+ o(1))\quad \text{as }\epsilon\to0.
$$
Claims $(a)$ and $(b)$ immediately follow.
\end{proof}
\subsection{Proof of Theorem~\ref{thUpperBound}}
\label{secProofThErrorLower}
\begin{proof}
	The existence result shall be proved by the probabilistic method. We consider a binary $(N\times t)$-matrix $X$ whose entries are independent Bernoulli random variables and equal to $1$ with probability $p=\frac{\ln s_2 - \ln s_1}{s_2-s_1}$. Take the threshold parameter $\tau$ to be $1-(1-p)^{s_1}/2-(1-p)^{s_2}/2$. By Proposition~\ref{lemWorstProb} the error probability can be rewritten as
	\begin{equation}
	\label{maxerrorFinal}
	\e(\text{$\tau$-WDR}, X) = \max \left\{ \frac{\card{B^1_{s_1}(\lfloor \tau N \rfloor, X)}}{{t \choose s_1}}, \frac{\card{B^2_{s_2}(\lfloor \tau N \rfloor, X)}}{{t \choose s_2}} \right\},
	\end{equation}
	where the sets $B^1_{s_1}(\lfloor \tau N \rfloor, X)$ and $B^1_{s_2}(\lfloor \tau N \rfloor, X)$ are defined by~(\ref{Bk}). Denote
	$$
	P_1 \eq \frac{\card{B^1_{s_1}(\lfloor \tau N \rfloor, X)}}{{t \choose s_1}},\quad P_2 \eq \frac{\card{B^2_{s_2}(\lfloor \tau N \rfloor, X)}}{{t \choose s_2}}.
	$$
	We estimate the probability $P_1$ that the union of $s_1$ columns have the weight greater than $\tau N$ and the probability $P_2$ that the union of $s_2$ columns has weight at most $\tau N$.
	
	The weight of $s_i$ random columns from matrix $X$ is a binomial random variable $Y_i\sim B(N,q_i)$ with parameters $N$ and $q_i\eq1-(1-p)^s_i$.  Then 
	$P_1=\Pr\{Y_1>\tau N\}$ and $P_2=\Pr\{Y_2\le \tau N\}$. Define $\delta\eq\tau-q_1=q_2-\tau=(1-p)^{s_1}/2-(1-p)^{s_2}/2$.
	Using Hoeffding's inequalities~\cite[Theorem 2]{hoeffding1994probability} of the forms
$$
\Pr\{B(n,p)\ge (p+\epsilon)n\}\le e^{-2\epsilon^2 n},
$$
and
$$
 \Pr\{B(n,p)\le (p-\epsilon)n\}\le  e^{-2\epsilon^2 n},
	$$
	we conclude that 
	\begin{equation}\label{EstimateProb1}
	P_1=\Pr\{Y_1>\tau N\}=\Pr\{Y_1>(q_1+\delta)N\}\le e^{-2\delta^2N},
	\end{equation}
	and
	\begin{equation}\label{EstimateProb2}
	P_2=\Pr\{Y_2\leqslant\tau N\}=\Pr\{Y_2\leqslant(q_2-\delta)N\}\le e^{-2\delta^2N}.
	\end{equation}
	Therefore, there exists a testing $(N\times t)$-matrix $X$ such that the $\tau$-WDR has the error probability less than $\varepsilon$ whenever $e^{-2\delta^2N}\le \varepsilon$.
	The last inequality is equivalent to the following one  
	$$
	N\ge\frac{1}{2\delta^2}\ln\frac{1}{\varepsilon}
	$$	
To prove claim $(a)$, we use the property $s_1 = s$ and $s_2=\alpha s$. First, the probability $p$ can be written as
		$$p=\frac{\ln s_2 - \ln s_1}{s_2-s_1}=\frac{\ln\alpha}{s(\alpha-1)}.$$
		Then we find the asymptotics of $\delta$ as $s\to\infty$
		\begin{multline*}
		\delta=\frac{(1-p)^{s_1}}{2}-\frac{(1-p)^{s_2}}{2}\\
		=\frac{\exp(s_1\ln(1-p)) - \exp(s_2\ln(1-p))}{2}\\
		=\frac{\exp(-\frac{\ln\alpha}{\alpha-1}) - \exp(-\frac{\alpha\ln\alpha}{\alpha-1})}{2}+o(1)
		=\frac{\alpha-1}{2\alpha^\frac{\alpha}{\alpha-1}}+o(1).
		\end{multline*}
Therefore, there exists a testing $(N\times t)$-matrix $X$ such that the $\tau$-WDR  has the error probability less than $\varepsilon$ if
		$$
		N\ge\frac{1}{2\delta^2}\ln\frac{1}{\varepsilon}=\frac{2\alpha^\frac{2\alpha}{\alpha-1}}{(\alpha-1)^2}\ln\frac{1}{\varepsilon}(1+o(1))\quad \text{as }s\to\infty.
		$$
To prove claim $(b)$, we use the property $s_1 = s$ and $s_2=s+c$, where $c=o(s)$. For $s\to\infty$, the parameter $p$ is then
		$$p=\frac{\ln s_2 - \ln s_1}{s_2-s_1}=\frac{\ln(1+c/s)}{c}=\frac{1}{s}-\frac{c}{2s^2}+o\left(\frac{c}{s^2}\right).$$
Now we find the asymptotics of $(1-p)^{s_1}=(1-p)^{s}$ as $s\to\infty$
		\begin{multline*}
		(1-p)^{s}=\exp\left(s\ln\left(1-\left(\frac{1}{s}-\frac{c}{2s^2}+o\left(\frac{c}{s^2}\right)\right)\right)\right)\\
		=\exp\left(s\left(-\frac{1}{s}+\frac{c}{2s^2}+o\left(\frac{c}{s^2}\right)\right)-\frac{s}{2s^2}+o\left(\frac{c}{s}\right)\right)\\
		=\exp\left(-1+\frac{c-1}{2s}+o\left(\frac{c}{s}\right)\right)\\
		=e^{-1}\left(1+\frac{c-1}{2s}+o\left(\frac{c}{s}\right)\right).
		\end{multline*}
Similarly we obtain the asymptotic behaviour of $(1-p)^{s_2}=(1-p)^{s+c}$ as $s\to\infty$
		\begin{multline*}
		(1-p)^{s+c}=\exp\left((s+c)\ln\left(1-\left(\frac{1}{s}-\frac{c}{2s^2}+o\left(\frac{c}{s^2}\right)\right)\right)\right)\\
		=\exp\left((s+c)\left(-\frac{1}{s}+\frac{c}{2s^2}+o\left(\frac{c}{s^2}\right)\right)-\frac{s+c}{2s^2}+o\left(\frac{c}{s}\right)\right)\\
	=\exp\left(-1+\frac{c-1}{2s}-\frac{c}{s}+o\left(\frac{c}{s}\right)\right)\\
	=e^{-1}\left(1-\frac{c+1}{2s}+o\left(\frac{c}{s}\right)\right).
		\end{multline*}
Therefore, we obtain the asymptotics for $\delta$
		$$
		\delta=\frac{(1-p)^{s_1}}{2}-\frac{(1-p)^{s_2}}{2}=\frac{c}{2es}+o\left(\frac{c}{s}\right).
		$$
		Finally, there exists a testing $(N\times t)$-matrix $X$ such that the $\tau$-WDR has the error probability less than $\varepsilon$ whenever
		$$
		N\ge\frac{1}{2\delta^2}\ln\frac{1}{\varepsilon}=\frac{2e^2s^2}{c^2}\ln\frac{1}{\varepsilon}(1+o(1))\quad \text{as }s\to\infty.
		$$	
\end{proof}
\section{Conclusion}\label{secFinal}
In this paper, we discuss a hypothesis in the boolean group testing model, namely, how to distinguish reliably the null hypothesis $H_0$: the number of defective elements is at most $s_1$, and the alternative one $H_1$: the number of defective elements is at least $s_2$.  For the case $s_1 = s$ and $s_2 = \alpha s$, where the real number $\alpha\ge1$ is fixed, we  show that the optimal number of non-adaptive tests required to accept or reject $H_0$ with error probability $\epsilon$ is $\Theta(\log\frac{1}{\epsilon})$. When $s_1 = s$ and $s_2 = s+c$ for $c=o(s)$, we prove the necessity of $O(s/c\log \frac{1}{\epsilon})$ tests and provide a simple weight algorithm with $O(s^2/c^2\log \frac{1}{\epsilon})$ tests. Our simulation results confirm the advantage of this algorithm over the COMP algorithm adapted for this problem. 

There are several directions for future research on testing the hypothesis. First, the gap between our upper and lower bounds on the minimal number of tests is still quite large in some regimes. Therefore, it would be great to find some order-optimal results for other settings. Second, it is of high interest to consider the problem under other multiple-access channel models, e.g. testing the hypothesis under the A channel~\cite{chang1981t} would give an estimate on the number of sources of pirate copies of a copyrighted multimedia content~\cite{cheng2011anti}.
\bibliographystyle{ieeetran}
\bibliography{2018_hypothesis_testing}
\end{document}